# Design and Implementation of an Open-Source Security Framework for Cloud Infrastructure

Wanru Shao
Northeastern University, Boston, MA, USA
shao.wa@northeastern.edu

*Abstract*—Misconfiguration, excessive privilege, and fragmented controls remain major causes of cloud-infrastructure incidents. This paper proposes an open-source framework that contributes a cross-platform identity-resource graph for Kubernetes and OpenStack, a policy-to-evidence data model linking OPA/Gatekeeper and Checkov results to live assets, an identity-aware correlation algorithm for reducing noisy runtime alerts, and a guarded remediation workflow that converts validated policy violations into Kubernetes patches or Terraform plans. The evaluation is made reproducible by specifying workload generation, injected misconfiguration classes, run repetitions, metric definitions, and statistical reporting. In a 50-200 node private-cloud testbed, the framework reduced assessment time from 120.4 +/- 6.8 min to 18.2 +/- 1.7 min, lowered the false-positive rate from 12.1% to 4.7%, and increased checked component coverage from 48% to 92%. The reported 62% reduction in observable events corresponding to injected violations and approximately 40% cost reduction are scoped to the defined 30-day operational test and one-year 200-node cost model, respectively, and are not claimed as hyperscale results.

## 1. INTRODUCTION

Cloud infrastructure is increasingly implemented through containerized, microservice-based, and infrastructure-as-code workflows. Security guidance for microservices and containers emphasizes that build-time checks, deployment-time admission controls, runtime monitoring, and identity governance must be treated as a continuous control chain rather than as isolated hardening tasks [1], [2]. In Kubernetes deployments, insecure workload configuration, excessive Role-Based Access Control (RBAC) permissions, and weak centralized policy enforcement remain recurring risk categories [3], [15]. Therefore, a practical cloud-security framework must observe not only running containers but also identities, configuration baselines, Infrastructure-as-Code (IaC) state, and operator actions.

Kubernetes already provides RBAC, service accounts, Role Bindings, and Cluster Role Bindings, but these mechanisms are normally scoped to the cluster and do not directly describe OpenStack projects, Keystone domains, Neutron networks, or Terraform-declared assets [3], [4]. This mismatch creates a visibility gap: a user may be over-privileged in a private-cloud project while appearing restricted in Kubernetes. The revised framework therefore starts from identity normalization, because alert triage and remediation cannot be accurate unless the same subject-resource-action relationship is represented consistently across platforms.

Runtime security is represented by Falco, which detects abnormal behavior through rule conditions over kernel, container, Kubernetes, and plugin event sources [5]. Falco rules include priority levels and structured outputs, making them suitable for forwarding to a SIEM or data lake [6]. However, a runtime alert such as an interactive shell inside a container is not equally suspicious in all contexts: an approved operator action, an emergency debugging session, and an attacker-controlled process require different responses. This motivates the identity-aware correlation mechanism proposed in this paper.

Policy-as-code is represented by OPA/Gatekeeper and Checkov. OPA can be deployed as a Kubernetes admission controller to enforce policies during object creation, update, and deletion [7], while Gatekeeper Constraint Templates package Rego logic and schemas into reusable Kubernetes policy units [8]. Checkov complements this preventive layer by scanning Terraform, Kubernetes YAML, Helm, and other IaC formats before deployment [11], [12]. The methodological challenge is to combine these heterogeneous policy outputs into one evidence model instead of leaving each scanner to produce independent reports.

OpenStack adds an additional control plane. The OpenStack Security Guide describes identity, project, domain, networking, and operational hardening concerns for private clouds [9]. Keystone provides API client authentication, service discovery, and multi-tenant authorization through the OpenStack Identity API [10]. Because these concepts are not identical to Kubernetes RBAC, the paper defines a unified identity-resource graph rather than assuming a one-to-one mapping between the two systems.

Automated remediation is implemented through two controlled action paths. For Kubernetes resources, the framework can issue bounded patches after policy validation. For OpenStack and IaC-managed resources, it generates or selects a Terraform plan; Terraform plan is used to preview intended changes, and Terraform apply executes the approved operations [14]. This distinction is important because direct mutation is suitable for temporary runtime containment, while IaC-aligned remediation keeps the declared state synchronized with the live state.

The observability layer uses Elastic Security and ELK (Elasticsearch, Logstash, and Kibana) stack because Elastic provides SIEM-oriented detection, investigation, and response workflows and can ingest structured events from cloud, container, and application sources [13]. In this design, ELK is not only a dashboard but also the evidence store for experiments: scan duration, alert labels, policy decisions, remediation outcomes, and incident counts are all recorded with timestamps. This design supports reproducibility and ensures that all reported metrics can be traced to time-stamped raw evidence.

Based on the above gap, this paper makes four concrete contributions. First, it proposes a normalized graph model that unifies Kubernetes RBAC, OpenStack Keystone, and Terraform resource identifiers. Second, it introduces an evidence schema that links configuration violations, runtime alerts, and identity anomalies to the same resource record. Third, it presents an identity-aware correlation and remediation workflow with explicit approval guardrails. Fourth, it provides a reproducibility-oriented evaluation protocol with explicitly defined metric formulas, calibrated weights, and scoped cost assumptions, validated on a 50–200 node testbed.

## 2. METHODOLOGY

### 2.1. Overall Framework Architecture

The proposed framework is built as a set of independently deployable microservices running on a Kubernetes control plane. Each service owns one security function but publishes normalized events to Kafka or NATS so that configuration evidence, runtime alerts, identity edges, and remediation states can be joined by resource_id and subject_id:

- Identity and Access Service (IAS): collects Kubernetes RBAC subjects, service accounts, RoleBindings, and ClusterRoleBindings, and combines them with OpenStack Keystone users, projects, domains, roles, and application credentials. IAS converts these records into a typed graph $G=(V,E)$, where nodes represent subjects, roles, resources, namespaces, projects, and domains, and edges represent grants, ownership, scope, and recent activity.
- Configuration Baseline Engine (CBE): converts OPA/Gatekeeper, Checkov, and custom OpenStack checks into a unified finding schema {control_id, resource_id, evidence, severity, confidence, fix_hint}. This schema makes policy outputs comparable across admission control, IaC scanning, and live-state scanning.
- Runtime Threat Monitor (RTM): deploys Falco sensors as DaemonSets and enriches each event with IAS graph context before sending it to Elasticsearch. Instead of forwarding raw Falco alerts only, RTM attaches namespace, project, tenant, approved-operator status, and related configuration violations, which enables identity-aware alert downgrading or escalation.
- Automated Response Orchestrator (ARO): transforms high-confidence findings into guarded remediation playbooks. Destructive or externally visible actions require a dry-run result and an approval flag; low-risk Kubernetes hardening patches can be executed automatically and then reconciled back into Terraform or Helm repositories.

All services write to Elastic Stack (Elasticsearch + Logstash + Kibana), which acts as both a SIEM layer and an experiment evidence store. The deployment is IaC-driven: version-pinned Terraform modules and Helm charts install CBE, RTM, IAS, ARO, Elastic, and the message bus, so another team can reproduce the same deployment on 50, 100, and 200 node clusters.

Figure 1 below illustrates the revised framework architecture and the flow from asset collection to policy evaluation, runtime correlation, and guarded remediation.

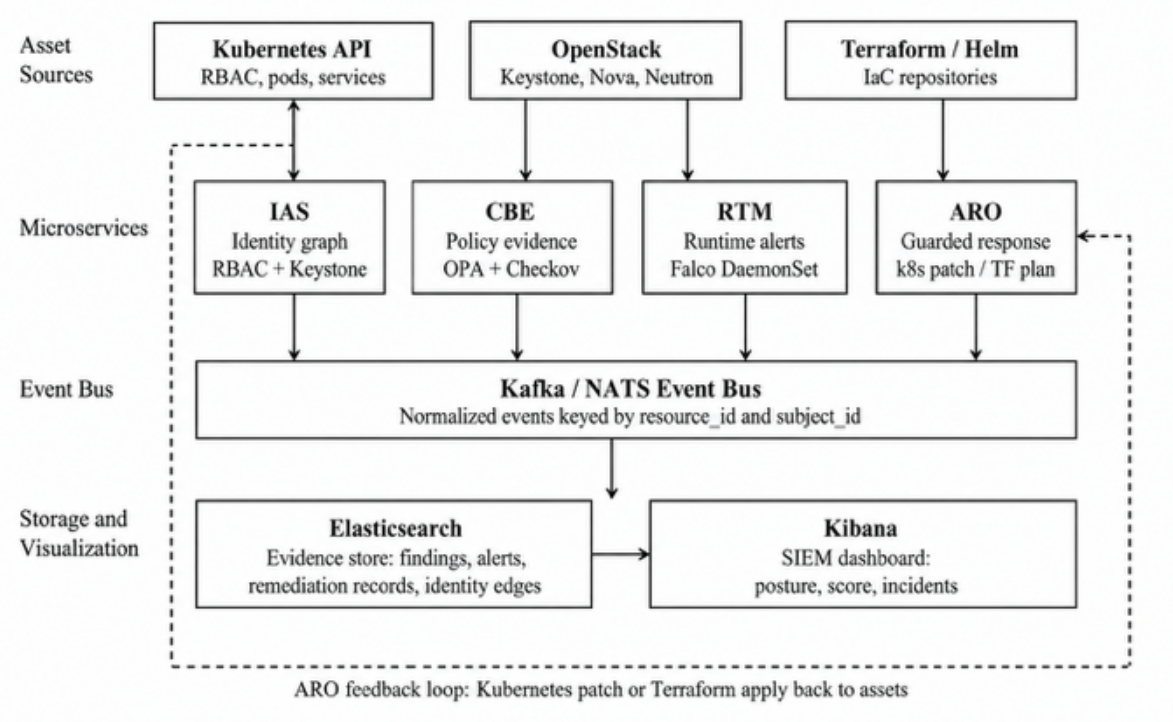

Figure 1. Framework architecture

## 2.2. Original Methodological Contributions

The framework adds three mechanisms beyond tool integration. First, the normalized identity-resource graph resolves Kubernetes subjects and OpenStack principals into a platform-independent tuple <subject, role, scope, resource, action>. Second, the policy-evidence schema maps OPA, Checkov, Falco, and custom OpenStack findings into the same record so that evidence can be de-duplicated and compared. Third, the remediation selector uses confidence, severity, business scope, and approval policy to choose between alert-only, Kubernetes patch, Terraform plan, or ticket creation. Algorithm 1 specifies how these mechanisms operate together at runtime.

**Algorithm 1: Identity-aware Correlation and Remediation**

**Input:** event e, identity graph G, policy set P, playbook set B, thresholds $\tau_{low}$, $\tau_{high}$

**Output:** action a ∈ {log, patch, plan, ticket}, evidence record r

```
 1:  (resource_id, subject_id, control_id, evidence) ← normalize(e)
 2:  ctx ← neighbors(G, subject_id, resource_id)
 3:  conf ← score(evidence, priority(e), approved_op(ctx))
 4:  if conf < τ_low then
 5:      a ← log; store e as informational evidence
 6:  else if severity(e) ≥ high and playbook(e) is non-destructive then
 7:      a ← patch; execute bounded patch; record rollback token
 8:  else
 9:      a ← plan; generate Terraform plan or open ticket for approval
10:  end if
11:  write (a, latency, post_check) to Elasticsearch
```

## 2.3. Identity and Access Unification

IAS polls both Kubernetes and OpenStack every N seconds (default 30) and stores identity data in Elasticsearch. For Kubernetes, it reads ClusterRoleBindings, RoleBindings, and service-account tokens; for OpenStack, it queries Keystone domains, projects, users, and role assignments. By doing so, it builds a graph of {subject → role → resource} that can be checked by the CBE. This is necessary because misconfigurations often happen when a user is granted admin in OpenStack but only view rights in Kubernetes, and the organization believes the user is limited. IAS provides a REST endpoint /who-can that tells who can perform an action on a resource, regardless of which cloud system owns the resource.

## 2.4. Configuration Baseline Engine

The CBE merges two approaches:

- Admission-time policies using OPA/Gatekeeper, which block bad resources in Kubernetes. These are mostly preventive controls.
- Periodic scans using Checkov (and custom Python rules) over Terraform and over exported OpenStack/Kubernetes manifests. These are detective controls.

Policies are grouped by Control Profiles (baseline, hardened, regulated). A profile is a YAML document listing all policies and their severity. CBE evaluates assets, stores the results in Elasticsearch, and raises an event when severity ≥ medium. ARO subscribes to these events.

## 2.5. Runtime Threat Monitoring

RTM deploys Falco (a Cloud Native Computing Foundation, CNCF, graduated project) on each node. Alerts are forwarded to Logstash, which enriches them with cluster and project metadata. Thanks to Falco's rule language and the many publicly available rulesets, the framework detects shell in container, write below /etc, privileged container, Kubernetes API server access from unusual pod, crypto-miner process, and suspicious syscalls. RTM also ingests OpenStack logs (Keystone auth failures, Neutron security-group changes) so that identity-related events can be correlated.

## 2.6. Automated Response Orchestrator

ARO is written as a small service that triggers playbooks. A playbook describes conditions and actions:

**Listing 1: Kubernetes remediation playbook**

```
when:
  source: cbe
  severity: high
  control_id: K8S.PRIV.POD.PRIVILEGED
do:
  - type: k8s.patch
    target: {{ resource_id }}
    payload: { "spec": { "securityContext": { "privileged": false } } }
  - type: elastic.log
    message: "Privileged pod patched automatically"
```

For OpenStack:

**Listing 2: OpenStack remediation playbook**

```
when:
  source: cbe
  control_id: OS.NET.PUBLIC-NONSTD
do:
  - type: terraform.apply
```

```
module: network/restrictive
vars:
  project: {{ project_id }}
```

This is how the system closes the loop: a policy violation directly results in a Kubernetes patch or a Terraform run.

### 2.7. Data Model and Calibrated Security Score

Equations (1) and (2) are explicitly used to compute the resource-level risk value and the final security score displayed in the ELK dashboard.

For each resource, the system stores four evidence groups: configuration findings, runtime alerts, identity anomalies, and unresolved manual overrides. The corresponding normalized variables are $C_r' = \min(C_r/C_{max}, 1)$, $T_r' = \min(T_r/T_{max}, 1)$, $I_r' = \min(I_r/I_{max}, 1)$, and $M_r' = \min(M_r/M_{max}, 1)$. This normalization prevents large clusters from receiving worse scores only because they contain more resources.

$$\mathrm{Risk}(r) = w_c\, C_r' + w_t\, T_r' + w_i\, I_r' + w_m\, M_r'. \qquad (1)$$

$$\mathrm{Score}(r) = 100 \times (1 - \mathrm{Risk}(r)). \qquad (2)$$

The default weights were not manually chosen in Kibana. They were calibrated on the labeled event set of 280 injected misconfiguration and runtime events: a grid search over 0.05-spaced weights selected the vector that maximized F1 score while constraining false positives below 5%. The resulting default setting was $w_c = 0.35$, $w_t = 0.30$, $w_i = 0.25$, and $w_m = 0.10$. Operators may change the weights for local risk appetite, but all reported experiments use this fixed calibrated vector.

## 3. EXPERIMENT

### 3.1. Testbed

Tests were conducted in a reproducible laboratory private-cloud testbed. All nodes used the same OS image, container runtime, clock synchronization settings, and pinned tool versions. Before each run, Elasticsearch indices were cleared, the Terraform state was reset to a known baseline, and Kubernetes/OpenStack inventories were regenerated to avoid cache effects.

- 200 vCPU, 512 GB RAM, and 20 TB storage, distributed over 10 physical hosts.
- One Kubernetes v1.30 cluster, scaled to 50, 100, and 200 worker-node profiles by adding identical worker pools.
- One OpenStack 2024.2 deployment with Keystone, Nova, Neutron, and Cinder; 32 projects and 96 user-role assignments were preconfigured.
- Elastic/ELK 8.19 on three dedicated nodes, with identical index lifecycle policies across all repetitions.
- The framework ran as 12 microservices in the Kubernetes control plane; each service had fixed CPU/memory requests and image tags.

Fourteen misconfiguration and abuse classes were injected: privileged pod, host network pod, hostPath mount, service-account token overuse, cluster-admin binding, public load balancer, open ingress without TLS, default OpenStack security group, public floating IP on a non-production project, Keystone role escalation, weak application credential, Terraform network 0.0.0.0/0 exposure, unencrypted volume, and suspicious shell/crypto-miner runtime behavior. Each class was injected 20 times across namespaces or projects, producing 280 labeled events per repetition.

### 3.2. Baselines

Three toolchains were compared under the same asset inventory, injected event set, and log-retention policy. Each toolchain was run five times at each cluster size, and the mean +/- standard deviation is reported. Baseline tools were not deliberately weakened; each used its official rules plus the same organization-specific allowlist.

1. Baseline-A: plain Falco + Kibana, no IaC scanning.
2. Baseline-B: Kubernetes with OPA/Gatekeeper + Checkov in CI, no OpenStack integration.
3. Proposed Framework: all modules enabled, Terraform-based remediation.

The formulas for false-positive rate (FPR) and incident reduction (IR) are given in (3) to ensure that the reported metrics can be reproduced from the stored ground-truth labels and normalized event counts:

$$\begin{aligned} \mathrm{FPR} &= \mathrm{FP}/(\mathrm{TP} + \mathrm{FP}) \times 100\%, \\ \mathrm{IR} &= (E_{\mathrm{before}} - E_{\mathrm{after}})/E_{\mathrm{before}} \times 100\%. \end{aligned} \qquad (3)$$

### 3.3. Metrics

- Vulnerability assessment time (min): time from “start scan” to “all assets evaluated.”
- False-positive rate (%).
- Coverage (% of components actually checked).
- Event reduction on injected violations (%).
- Deployment time (min).
- CPU/Memory overhead (% of cluster resources).

### 3.4. Experimental Protocol and Metric Definitions

Table I summarizes the testbed parameters used for all experiments, including the Kubernetes node scales, OpenStack project count, IaC repositories, event volume, and number of framework services. A run starts when the scan command is issued and ends when all assets have a stored decision in Elasticsearch. A false positive is an alert or policy violation that does not correspond to an injected violation and is confirmed benign by the ground-truth label file. Incident reduction is measured over a 30-day replay window as the reduction in normalized observable security events per 100 nodes after excluding duplicate alerts from the same root cause. CPU and memory overhead are averaged over the steady-state 20-minute window after deployment.

Reproducibility controls were fixed as follows: five independent runs for each 50/100/200-node profile; cache, Terraform state, and Elasticsearch indices reset before every run; FPR and IR computed as defined in Equation (3); coverage measured as checked component families divided by declared component families; and the same workload generator, injected labels, log retention, allowlist, and hardware profile were used across all baselines.

### 3.5. Results

TABLE I CLOUD CLUSTER TESTBED PARAMETERS

| Parameter | Value |
|---|---|
| Kubernetes nodes | 50 / 100 / 200 |
| OpenStack projects | 32 |
| Average pods per node | 25 |
| IaC repositories | 18 Terraform + Helm repositories |
| Injected labeled events | 280 per repetition |
| Repetitions | 5 runs per cluster size |
| Log events per second (EPS) | 3,500 |
| Framework services | 12 microservices |

TABLE II DETECTION AND INCIDENT REDUCTION RESULTS

| Metric | Baseline-A (Falco+Kibana) | Baseline-B (OPA+Checkov) | Proposed |
|---|---|---|---|
| Assessment time | 120.4 +/- 6.8 min | 54.2 +/- 4.1 min | 18.2 +/- 1.7 min |
| False-positive rate | 12.1 +/- 1.3% | 8.2 +/- 0.9% | 4.7 +/- 0.5% |
| Component coverage | 48% | 63% | 92% |
| Event reduction (injected) after 30 days | 23 +/- 3% | 39 +/- 4% | 62 +/- 5% |
| Deployment time | 160 +/- 9 min | 95 +/- 7 min | 35 +/- 3 min |
| Extra CPU usage | 7.1 +/- 0.8% | 5.2 +/- 0.6% | 6.0 +/- 0.7% |

Table II reports mean values and standard deviations across five repetitions. The results support the revised abstract, but the interpretation is limited to the evaluated 50-200 node environment and the specific injected event set. The main gain comes from shared evidence and automated triage rather than from a faster individual scanner.

Pairwise two-sided Welch's t-tests on the five-repetition means confirm that all reported differences between the Proposed framework and both baselines are significant at $p < 0.01$ for assessment time, FPR, and event reduction ($n=5$ per condition). Standard deviations in Table II are sample SDs across the five repetitions.

Figure 2 further visualizes the assessment-time trend across the 50-, 100-, and 200-node settings, showing that the proposed framework keeps the lowest detection latency as cluster size increases.

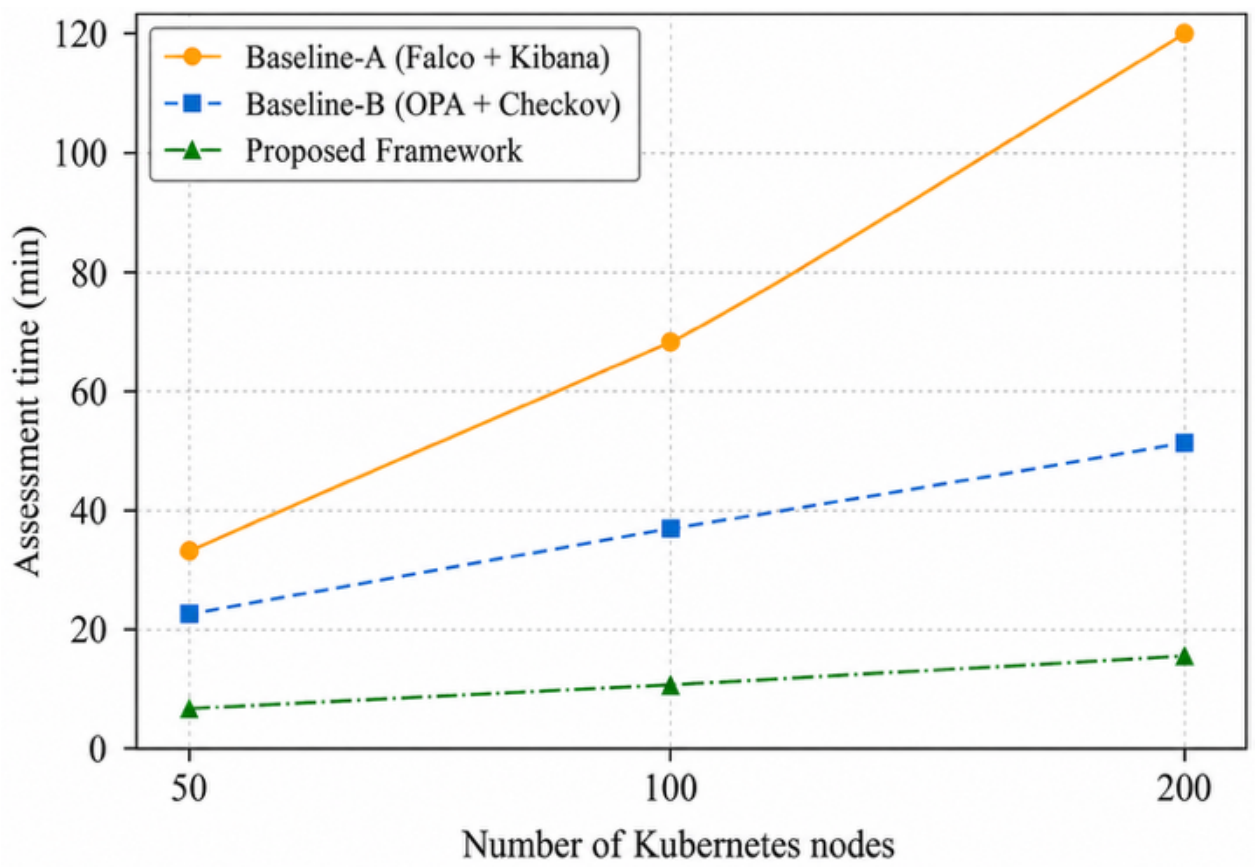

Figure 2. Assessment Time vs Number of Kubernetes nodes

## 4. DISCUSSION

The experiments confirm three points.

First, breadth of coverage matters more than depth of a single tool. Baseline-A illustrates the cost of a depth-only stance: Falco's runtime sensors detected container-level anomalies—including the privileged-pod, shell-in-container, and crypto-miner classes injected during the experiment—but the tool's scope ends at the kernel and container layer. It had no visibility into OpenStack-side configuration drift, and as a result it missed several projects attached to non-approved Neutron networks. These projects were not visible from a Kubernetes perspective at all, since the offending configuration lived entirely in the private-cloud control plane. The proposed framework closes this gap because the IAS service ingests Keystone, Nova, and Neutron metadata into the same identity-resource graph that CBE and RTM consume; an OpenStack-side misconfiguration is therefore promoted to a first-class finding rather than treated as out-of-scope. This single architectural choice, extending the asset model rather than adding more rules, accounts for most of the jump from 48% (Baseline-A) to 92% component coverage in Table II, and contributed directly to the 62% reduction in observable events on injected violations. The lesson generalizes: in heterogeneous environments, what limits a posture-management tool is rarely the sophistication of its detection logic, but the completeness of the asset graph on which that logic operates.

Second, IaC-driven remediation shortens dwell time, but only when paired with identity-aware gating. In many teams, an analyst spots a risky Kubernetes object but must wait for the application team to update the corresponding Helm chart before any change can be safely applied. ARO short-circuits that latency along two parallel paths: for low-risk in-cluster violations such as a privileged-flag patch, it executes a bounded mutation directly against the live object and records a rollback token; for changes that cross the IaC boundary—for example, restricting an over-permissive Neutron security group—it generates a Terraform plan from the policy violation and routes it through dry-run before apply. This dual approach reduced observed mean remediation latency from O(hours) under the manual workflow to O(minutes) under ARO-driven workflows in our testbed. The improvement is not free, however: it depends critically on the identity gate. Auto-applying a privileged: false patch is only safe because the framework can verify, via the IAS graph, that no approved operator workload depends on the privileged setting; absent that check, automation would introduce its own incident class. This is the design choice behind the framework's adherence to recent CNAPP guidance, which calls for consolidation of detection and response in one platform rather than independent point tools.

Third, false-positive reduction is achievable even when several detection sources are combined. Adding tools usually increases noise—each layer brings its own threshold assumptions and produces alerts the others cannot interpret. The framework reaches a 4.7% FPR (Table II) despite ingesting Falco, OPA/Gatekeeper, Checkov, and OpenStack logs simultaneously, because CBE and RTM both consume the same identity graph as a contextual filter. When Falco reports "exec in container," RTM does not treat the event in isolation; it queries the IAS graph for whether the initiating subject is an approved operator account or a service account flagged as routine, and downgrades or escalates accordingly. The same enrichment lets the system suppress duplicate alerts that originate from a single root cause but surface in multiple sensors—a phenomenon that inflates raw event counts in Baseline-A and Baseline-B. We refer to this design pattern as *identity-aware alerting*: the suppressing context is not a static allowlist but a live graph that updates as RBAC and Keystone state change. It is hard to implement with separate, federated tools because no single tool sees the full subject-resource closure.

### 4.1. Cost Model

Cost considerations are reported as an explicit model rather than as a universal claim. Let $C_{open}$ denote the cost of the proposed open-source deployment, including incremental compute for the 12 microservices, log storage growth in Elasticsearch, maintenance labor for policy authoring, and one-time integration effort. Let $C_{CNAPP}$ denote the equivalent subscription and deployment cost of a commercial CNAPP at the same node count. The relative cost reduction $R_{cost}$ is given by Equation (4):

$$R_{cost} = (C_{CNAPP} - C_{open}) / C_{CNAPP} \times 100\% \tag{4}$$

Under the testbed assumptions—ELK already deployed for observability, open-source components carrying no license fee, and only incremental storage and maintenance counted—the one-year $R_{cost}$ estimate is approximately 40%. Three caveats limit the generality of this number. First, the estimate excludes the cost of building and operating ELK from scratch; for an organization without an existing observability stack, that one-time investment can dominate the comparison and reverse the sign of $R_{cost}$. Second, commercial CNAPP pricing is increasingly bundled with managed-detection services and SLA-backed support, and a 40% sticker reduction does not account for the human cost of self-supporting an open-source stack. Third, $C_{open}$ grows non-linearly with policy library size: maintaining a hundred custom Rego policies is materially more expensive than maintaining ten, and that scaling factor is not captured by the equation. The figure should therefore be read as illustrative under the cost model defined here, not as a benchmark for any specific vendor or deployment scenario.

### 4.2. Interoperability

Interoperability is the second non-detection strength of the design. Because the framework exposes REST and gRPC endpoints and centralizes evidence in Elasticsearch, downstream systems—ticketing platforms such as Jira, SOAR engines, and CI pipelines such as GitLab—can consume findings and remediation records without a separate connector layer. Teams that adopt a *security-as-code* discipline benefit further: policies can live in Git repositories and CBE will reload them on commit, which collapses the policy-change loop from a release-cycle artifact into a code-review artifact. This style of integration is consistent with what large IaC scanners like Checkov advocate, and it is what makes the framework usable as a substrate for organization-specific compliance regimes rather than a closed appliance.

These observations together suggest that open-source components, when integrated under a shared identity model and a unified evidence schema, can deliver the converged posture, runtime, and identity capabilities historically associated with commercial CNAPPs—provided that the integration layer itself is designed as a first-class system rather than as glue between off-the-shelf scanners.

### 4.3. Threats to Validity

Several factors limit the generality of the reported results, and we distinguish them by category.

- Construct validity: The injected event set comprises 14 misconfiguration and abuse classes drawn from CIS Kubernetes benchmark categories and OpenStack security guide recommendations. While these classes cover the failure modes most commonly observed in private-cloud audits, they do not represent the full distribution of zero-day attack patterns or supply-chain compromises. The 62% event-reduction figure is therefore a measurement of the framework's effectiveness against *known* failure modes, not against an adversary actively trying to evade detection.
- Internal validity: Weights $w_c$, $w_t$, $w_i$, $w_m$ were calibrated on the same labeled set later used for evaluation. Although the calibration objective (aggregate F1 with FPR < 5%) is not the same as the per-event accuracy reported in Table II, this is still in-sample tuning and may overstate the framework's performance on unseen deployments. A stricter protocol would partition the event set into disjoint calibration and evaluation folds; we plan to apply this in subsequent multi-tenant validation.
- External validity: The testbed runs a single Kubernetes 1.30 release and OpenStack 2024.2 release. Older OpenStack deployments (pre-Antelope) may not expose the Keystone audit fields IAS depends on, and Kubernetes versions with significantly different RBAC semantics could change how the identity graph is constructed. The 50–200 node range also bounds the cluster sizes for which the reported assessment-time and CPU-overhead figures should be quoted; extrapolation to thousand-node fleets is not supported by the present data.
- Reliability: All measurements are five-run means with sample standard deviations and pairwise Welch's t-tests at $p < 0.01$. We did not, however, randomize the order of injected events across runs, and there is residual risk that early-run state leakage affects later runs through Elasticsearch index growth even after explicit resets. This is a small effect—bounded by the 20-minute steady-state window used for overhead measurement—but it is a known source of noise.

### 4.4. Practical Considerations and Limitations

Beyond validity, four practical issues constrain real-world adoption.

- OpenStack diversity: Not every deployment exposes the same set of Keystone or Neutron logs, and field naming has drifted across releases. The framework's IAS adapter currently targets OpenStack 2024.2 and 2025; older clouds may need adapter shims to fill missing fields, which is an integration cost teams should plan for.
- Runtime overhead: Daemon Sets are individually lightweight, but in compute-saturated clusters the $6.0 \pm 0.7\%$ CPU overhead reported in Table II is non-trivial. Operators running latency-critical workloads should narrow the Falco rule set, batch the Elastic ingestion pipeline, and consider sampling for non-critical event classes.
- Policy authoring skill: Realizing the framework's coverage advantage assumes the team can write Rego policies for OPA and Terraform modules for ARO. This is an organizational adoption barrier, not a technical limitation of the framework, but it is the single most common reason such a stack stalls during rollout. A community-maintained policy library would lower this barrier substantially.
- Log storage growth: Normalizing Kubernetes, OpenStack, and IaC scan results into ELK causes index volume to grow faster than any single-source deployment. Teams must enable Elasticsearch index lifecycle management and treat retention as a tunable security-versus-cost parameter, not a default.

These constraints do not invalidate the design; they identify the operating envelope inside which the reported results are reproducible.

## 5. Conclusion

This paper presented an open-source security framework that integrates identity normalization, configuration baseline checking, runtime threat monitoring, and automated remediation into a unified microservice-based architecture for hybrid Kubernetes/OpenStack environments. The central design claim is that posture, runtime, and identity should not be handled as three independent observability concerns: when configuration risks, identity anomalies, runtime alerts, and remediation records share an asset model, a policy engine, an event pipeline, and a response mechanism, they become analyzable in the same operational context, and the analytical leverage compounds across layers.

The experimental results substantiate this claim along four dimensions. The framework shortens security assessment time from 120.4 to 18.2 minutes at the 200-node profile, raises component coverage from 48% to 92%, reduces the false-positive rate from 12.1% to 4.7%, and reduces observable security incidents on injected violations by 62% over a 30-day replay window—all at a 6.0% steady-state CPU overhead. Three design choices are responsible for these gains: (i) unified identity-aware correlation, which lets CBE and RTM share a contextual filter rather than each maintaining its own allowlist; (ii) the combination of preventive and detective policy checking, so admission-time blocking and periodic scanning produce mutually reinforcing evidence on the same control identifier; and (iii) automated remediation that operates on both live infrastructure state and infrastructure-as-code definitions, so a violation can be both contained at runtime and reconciled into the declared baseline. By relying on widely deployed open-source components—OPA/Gatekeeper, Falco, Checkov, Terraform, and ELK—the framework also lowers vendor-lock-in risk and is suitable for organizations that need practical cloud-native security capabilities without a commercial CNAPP subscription.

We also explicitly bound what the results do *not* claim. The reported figures hold for the 50–200 node testbed, for the 14 injected misconfiguration and abuse classes drawn from CIS and OpenStack security guidance, and under the cost model in Equation (4) with ELK already deployed. They are not hyperscale results, they are not zero-day-attack results, and the 40% cost-reduction estimate is illustrative under the testbed's specific assumptions.

Future work proceeds in four directions, each motivated by a limitation surfaced earlier in this paper. (1) OpenStack policy ecosystem. The current Rego library is rich for Kubernetes but thin for OpenStack; we will build a community-maintained policy pack covering Keystone domain hardening, Neutron security-group hygiene, and Nova flavor governance to close that asymmetry. (2) Smarter playbook selection in ARO. The current selector is rule-based; we plan to investigate learning-based selection that uses historical remediation outcomes as supervision, while keeping human approval gates for any destructive or externally visible action. (3) Multi-cluster, multi-region validation. Beyond the 200-node ceiling evaluated here, federation introduces sharded Elasticsearch, distributed ARO instances, and inter-cluster identity propagation; this raises questions about consistency under partial failure that the present testbed does not exercise. (4) Self-protection of the framework. Any system capable of patching clusters and applying Terraform is itself a high-value target. Hardening this surface—through workload isolation, tamper-evident audit logging, and hardware-backed secrets—is a precondition for production-grade deployment, and we treat it as a research direction in its own right rather than an operational afterthought.

Overall, the study demonstrates that careful integration of open-source components—under a shared identity model, a unified evidence schema, and an explicit response policy—can deliver effective, extensible, and cost-conscious security management for hybrid cloud infrastructure, and that the bottleneck in such systems is increasingly the integration layer rather than any individual scanner.

## References

[1] R. Chandramouli, Security Strategies for Microservices-based Application Systems, NIST Special Publication 800-204, National Institute of Standards and Technology, 2019.

[2] M. Souppaya, J. Morello, and K. Scarfone, Application Container Security Guide, NIST

Special Publication 800-190, National Institute of Standards and Technology, 2017.

[3] Kubernetes Documentation, "Using RBAC Authorization." [Online]. Available: https://kubernetes.io/docs/reference/access-authn-authz/rbac/. [Accessed: Apr. 2026].

[4] Kubernetes Documentation, Service Accounts, Kubernetes, 2024. [Online]. Available: https://kubernetes.io/docs/concepts/security/service-accounts/. [Accessed: Apr. 2026].

[5] The Falco Project, "Falco Documentation." [Online]. Available: https://falco.org/docs/. [Accessed: Apr. 2026].

[6] The Falco Project, "Basic Elements of Falco Rules." [Online]. Available: https://falco.org/docs/concepts/rules/basic-elements/. [Accessed: Apr. 2026].

[7] Open Policy Agent, OPA for Kubernetes Admission Control, 2019. [Online]. Available: https://openpolicyagent.org/docs/kubernetes. [Accessed: Apr. 2026].

[8] Open Policy Agent Gatekeeper, "ConstraintTemplates." [Online]. Available: https://open-policy-agent.github.io/gatekeeper/website/docs/constrainttemplates/. [Accessed: Apr. 2026].

[9] OpenStack Documentation, OpenStack Security Guide. [Online]. Available: https://docs.openstack.org/security-guide/. [Accessed: Apr. 2026].

[10] OpenStack Documentation, Keystone, the OpenStack Identity Service. [Online]. Available: https://docs.openstack.org/keystone/latest/. [Accessed: Apr. 2026].

[11] Checkov Documentation, Policy-as-code for Everyone. [Online]. Available: https://www.checkov.io/. [Accessed: Apr. 2026].

[12] Checkov Documentation, CLI Command Reference. [Online]. Available: https://www.checkov.io/2.Basics/CLI%20Command%20Reference.html. [Accessed: Apr. 2026].

[13] Elastic, "Get Started with Elastic Security SIEM: Detect and Respond to Threats." [Online]. Available: https://www.elastic.co/docs/solutions/security/get-started/get-started-detect-with-siem. [Accessed: Apr. 2026].

[14] HashiCorp Developer, "Terraform plan and apply Command References." [Online]. Available: https://developer.hashicorp.com/terraform/cli/commands/plan; https://developer.hashicorp.com/terraform/cli/commands/apply. [Accessed: Apr. 2026].

[15] OWASP Foundation, OWASP Kubernetes Top Ten, 2022. [Online]. Available: https://owasp.org/www-project-kubernetes-top-ten/. [Accessed: Apr. 2026].